\documentclass[letterpaper,10pt,nofootinbib,aps,tightenlines,twocolumn]{revtex4}

\usepackage{amsmath,amsfonts,amssymb}
\usepackage{mathrsfs}
\usepackage{graphicx}

\usepackage[english]{babel}

\usepackage{booktabs}
\newcommand{\ra}[1]{\renewcommand{\arraystretch}{#1}}
\usepackage{appendix}

\usepackage{epstopdf}
 \usepackage{hyperref}

\begin{document}

\title{Freezing Out Early Dark Energy }

\author{Jannis Bielefeld$^1$, W. L. Kimmy Wu$^2$, Robert R. Caldwell$^1$, and Olivier Dor\'e$^{3,4}$}
\affiliation{$^1$Department of Physics \& Astronomy, Dartmouth College, 6127 Wilder Laboratory, Hanover, NH 03755 USA}
\affiliation{$^2$ Department of Physics, Stanford University, Stanford, CA 94305 USA}
\affiliation{$^3$ NASA Jet Propulsion Laboratory, California Institute of Technology, 4800 Oak Grove Drive, Pasadena, CA 91125 USA}
\affiliation{$^4$ California Institute of Technology, MC249-17, Pasadena, CA 91125 USA}

 \date{\today}

\begin{abstract}

A phenomenological model of dark energy that tracks the baryonic and cold dark matter at early times but resembles a cosmological constant at late times is explored. In the transition between these two regimes, the dark energy density drops rapidly as if it were a relic species that freezes out, during which time the equation of state peaks at $+1$. Such an adjustment in the dark energy density, as it shifts from scaling to potential-domination, could be the signature of a trigger mechanism that helps explain the late-time cosmic acceleration. We show that the non-negligible dark energy density at early times, and the subsequent peak in the equation of state at the transition, leave an imprint on the cosmic microwave background anisotropy pattern and the rate of growth of large scale structure. The model introduces two new parameters, consisting of the present-day equation of state and the redshift of the freeze-out transition. A Monte Carlo Markov Chain analysis of a ten-dimensional parameter space is performed to compare the model with pre-Planck cosmic microwave background, large scale structure and supernova data and measurements of the Hubble constant. We find that the transition described by this model could have taken place as late as a redshift $z\sim 400$. We explore the capability of future cosmic microwave background and weak lensing experiments to put tighter constraints on this model. The viability of this model may suggest new directions in dark-energy model building that address the coincidence problem.
 
\end{abstract}
\maketitle

\section{Introduction}

The recent domination of dark energy brings into question its past behavior -- where has it been hiding for the approximately ten billion years before the onset of cosmic acceleration? The answer for most theoretical models of dark energy, including the cosmological constant, is that the dark energy has been unnaturally tuned to comprise a negligible portion of the cosmic energy budget for most of history. Yet in hopes of explaining dark energy in more familiar terms, some theories propose that dark energy has been scaling with the radiation, e.g. Refs.~\cite{Wetterich:1987fm,Ratra:1987rm,Zlatev:1998tr,ArmendarizPicon:2000dh}, almost as if the dark energy was in thermal equilibrium with the cosmological fluid before undergoing a late-time phase transition \cite{Frieman:1991tu,Frieman:1995pm}. In this case, for most of history, dark energy contributed a constant fraction of the radiation, according to some equipartition. This scenario is sometimes referred to as early dark energy or early quintessence \cite{Doran:2001rw,Caldwell:2003vp,Amendola:2007yx} because the dark energy contributes a non-negligible fraction of the cosmic energy budget at early times. At very late times in this scenario, perhaps well into the matter-dominated era, a change in the dark energy is triggered, and the dark energy suddenly becomes potential-energy dominated \cite{Hebecker:2000au,Hebecker:2000zb}. While it remains an open problem in cosmology to find a plausible mechanism for such a trigger \cite{Copeland:2006wr}, we would like to consider more broadly what might be the observational signatures of a trigger should such a scenario be correct.

Nature may abhor a vacuum, but dark energy abhors couplings to standard forms of radiation and matter. A generic coupling between a light cosmic scalar field and the Standard Model, invoked to explain the scaling with radiation or to trigger a shift in the dark energy field, would ruin the ability of the scalar to accelerate the universe \cite{Carroll:1998zi}. However, little is known about dark matter so perhaps there is room for speculation about a ``dark sector" consisting of dark matter and dark energy (e.g. Refs.~\cite{Axenides:2004kb,Das:2005yj,Brookfield:2007au,Bean:2008ac,Cai:2009ht,He:2009pd}). We take this route and suppose that the dark energy may have something to do with dark matter (e.g. Refs.~\cite{Amendola:1999er,ArkaniHamed:2000tc,Mangano:2002gg}), such that the dark energy scales with the dark matter throughout most of cosmic history. However, we speculate that at some point the dark energy detaches or ``decouples'' and its energy density plummets until it reaches close to its asymptotic value and becomes potential-energy dominated. In borrowed language, we might say that the dark energy ``freezes out". In this paper we report on efforts to model this type of dark energy behavior in terms of a cosmic scalar field, quintessence, with a potential that yields the desired equation-of-state dynamics.  

Our model is ``early dark energy" (EDE) since it contributes a non-negligible fraction of the cosmic energy budget in the early stages of the matter-dominated era. However, the equation of state trajectory is very different from the canonical EDE models in which $w$ evolves monotonically from the radiation-like $1/3$ down towards the lambda-like $-1$ \cite{Doran:2006kp,Corasaniti:2002vg,Pettorino:2013ia}. Rather, $w$ starts at zero until it decouples and rises to $+1$, after which point it drops down close to $-1$. Although there have been general investigations of EDE and departures from the standard expansion history \cite{Hojjati:2013oya}, the consequences of a spike in the equation of state of EDE have not been widely explored.

In Sec.~\ref{sec:model} we present our model and explain its features, including a perturbation analysis. The data analysis and constraints are presented in Sec.~\ref{sec:constraints}. Parameter forecasts are given in Secs.~\ref{sec:lensing} and \ref{sec:tt}.

\section{Model}
\label{sec:model}

We propose a model in which the dark energy has a non-negligible contribution at early times but which drops off before coming to dominate. Having a non-vanishing contribution at early times can be achieved by requiring the equation of state to mimic the background component. In our case it suffices for the fluid to have matter behavior ($w\sim 0$). Dark energy domination requires $w$ to drop down below $w<-1/3$; in our case it drops down to $w\rightarrow -1$ by the present. Intermediate to these two regimes, the equation of state history is distinguished by a spike sending $w\rightarrow +1$. These three regimes determine the shape of the equation of state.

To build such a model, it is simplest to start with the energy density and postulate a dependence on the scale factor. We define
$\rho_{\rm DE}(a) =  \rho_{\rm DE}(a_0) (a_0/a)^{3 \gamma(a)}$ where
\begin{equation}
\gamma(a) = 1 + w_0 \left(\frac{1 + \frac{2}{\pi}\tan^{-1}k(x_c-x(a))}{1 +\tfrac{2}{\pi}\tan^{-1}k x_c}\right).
\end{equation}
In the above equation, $x \equiv \ln(a_0/a)$, $x_c = \ln(a_0/a_c)=\ln(1+z_c)$ where $a_c$ determines the dark-energy transition scale factor, $z_c$ is the corresponding redshift, and $k$ is a parameter to be determined. At early times, $\gamma \simeq 1$ in order for the dark energy to scale with matter.  At late times, after $x$ overtakes $x_c$, $\gamma$ approaches $1+w_0$. Using the above functional form for $\gamma$, we can derive the equation of state
\begin{equation}
w(a) = w_0 \left(\frac{1 + \frac{2}{\pi}\left[  \tan^{-1} k(x_c -x(a))-\frac{k x(a)}{1 + k^2(x(a)-x_c)^2} \right] }{1 +\tfrac{2}{\pi}\tan^{-1}k x_c}\right).
\end{equation}
The peak value of the equation of state is $w_p = w_0 (2/\pi)(\tan^{-1} k x_c - k x_c)/(1 +\tfrac{2}{\pi}\tan^{-1}k x_c)$. In this study, we fix $w_p = +1$. Consequently, a choice of $w_0$ determines the product $k x_c$. Since $x_c$ is also chosen, then $k$ is fixed. Hence, we have a two-parameter family of models, determined by $w_0$ and $x_c$.

The equation of state history described above can be achieved by a rolling scalar field. During the phase in which the scalar tracks the matter density, the shape of the potential is $V \propto \phi^{-n}$ where $n \simeq 6$ during the radiation-dominated era, but $n$ grows larger to approximate an exponential potential at the onset of matter domination. The spike in the equation of state is achieved by a sharp drop in the potential, as if the field was a ball rolling down a hill and off a cliff, whereafter the field is wedged in a narrow minimum at $V>0$. This potential is clearly contrived, and a more realistic model would require more physics; however our purpose is to focus on the spike in the equation of state, and so we leave for future work the task to develop a more realistic EDE model.

Other EDE studies have concentrated on equations of state that track the background behavior more closely -- in particular tracking the radiation component \cite{Corasaniti:2002vg}. Our approach differs in two ways. First, this model tracks pressureless matter at early times instead of radiation. Second, the equation of state decreases monotonically in most models. Instead, our equation of state peaks at $w=+1$ before the dark energy component transitions to $w\rightarrow -1$. This peak is responsible for the rapid loss of energy density when the dark energy component freezes out. In this respect this behavior resembles the freeze-out of a relic particle species even though dark energy does not couple to another component in order to lose its energy.
 
 \subsection{Background evolution}
\begin{figure}[htbp]
\includegraphics[width=\linewidth]{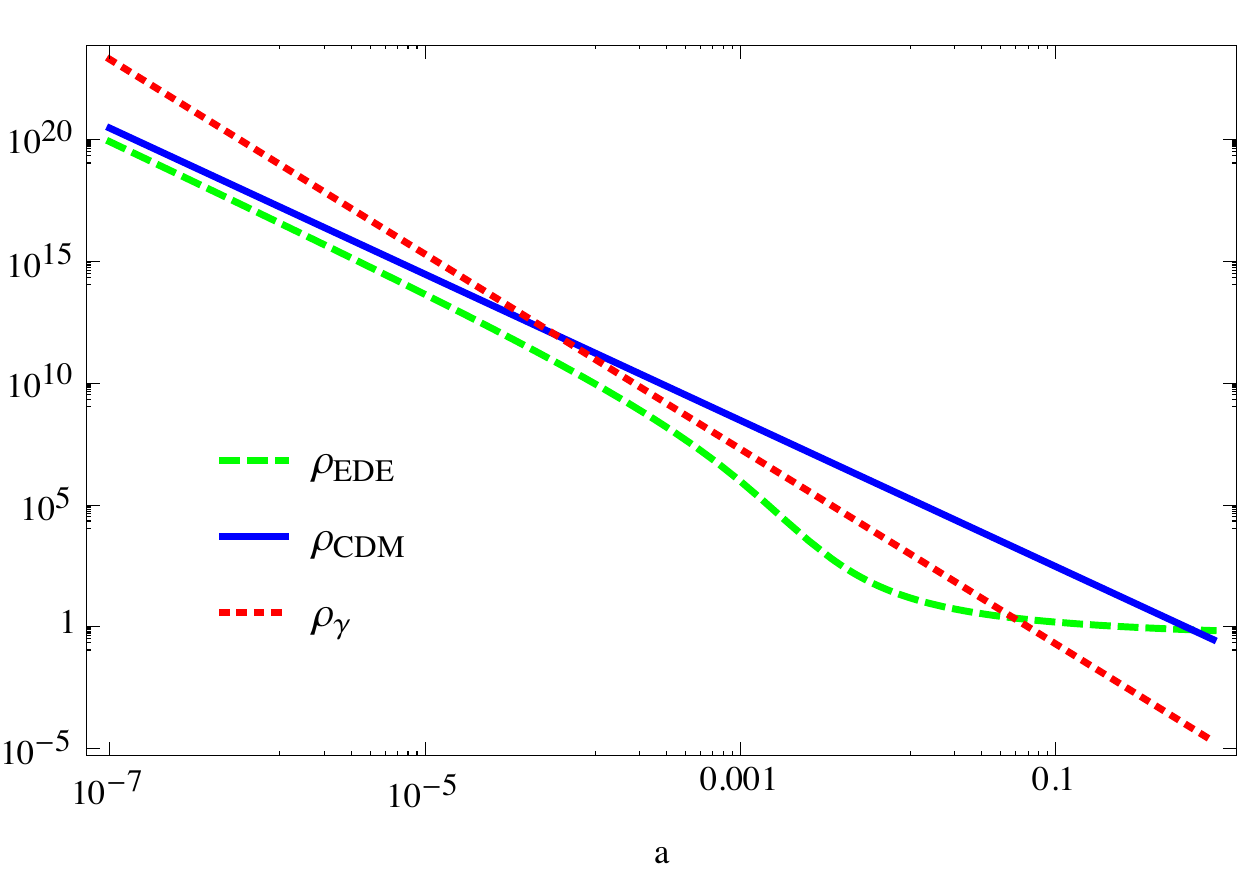}
\caption{Energy densities $\rho$ vs. scale factor $a$. At early times the dark energy component follows the matter content. At around $a=0.001$ its energy density drops significantly. It subsequently changes its behavior to $w \rightarrow -1$ (see Fig.~\ref{fig:fig3}). At late times dark energy dominates the expansion.}
\label{fig:fig1}
\end{figure}

\begin{figure}[htbp]
\includegraphics[width=\linewidth]{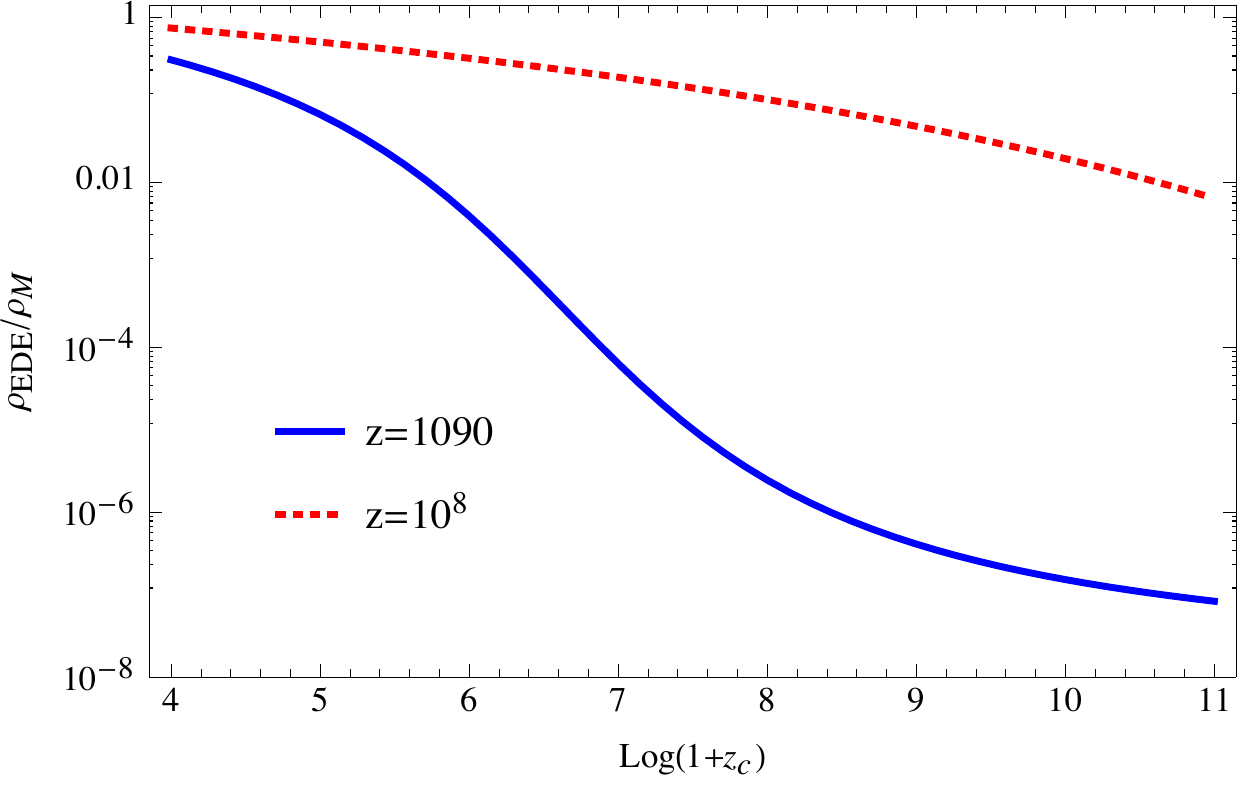}
\caption{The ratio of dark energy to matter at early times and at recombination is shown for a range of transition redshifts $z_c$. For a late transition as shown on the left end of the figure, the dark energy can have approximately the same energy density as matter in the early Universe, before dropping to the few percent level by the time the CMB is emitted. }
\label{fig:fig1b}
\end{figure}

\begin{figure}[htbp]
\includegraphics[width=\linewidth]{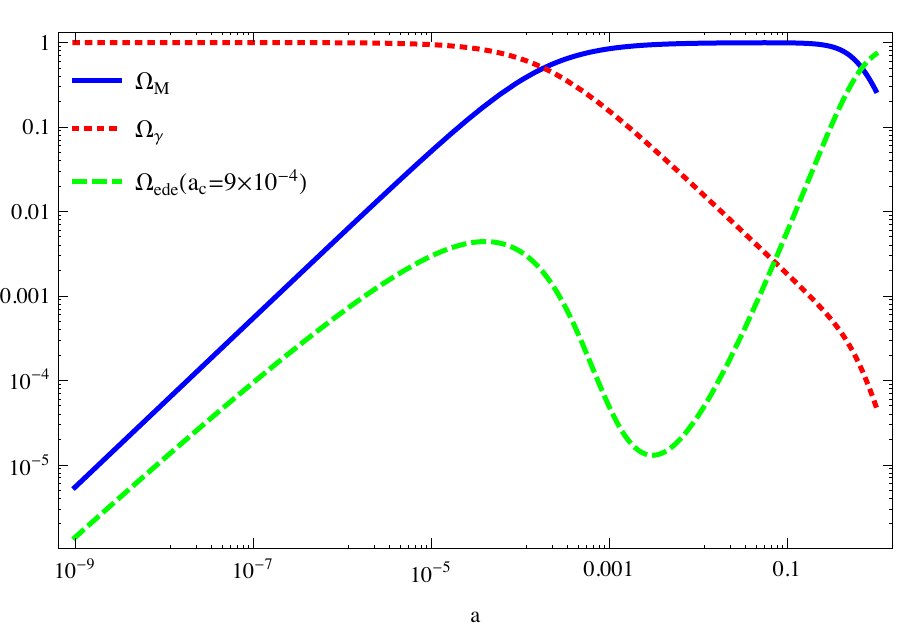}
\caption{Scaling behavior of the relative energy densities. Notice the maximum in $\Omega_{ede}$ at $a\sim 10^{-4}$. This leads to interesting deviations from the standard scenario of structure formation. (See Fig.~\ref{fig:perturbations}.)}
\label{fig:fig2}
\end{figure}

\begin{figure}[htbp]
\includegraphics[width=\linewidth]{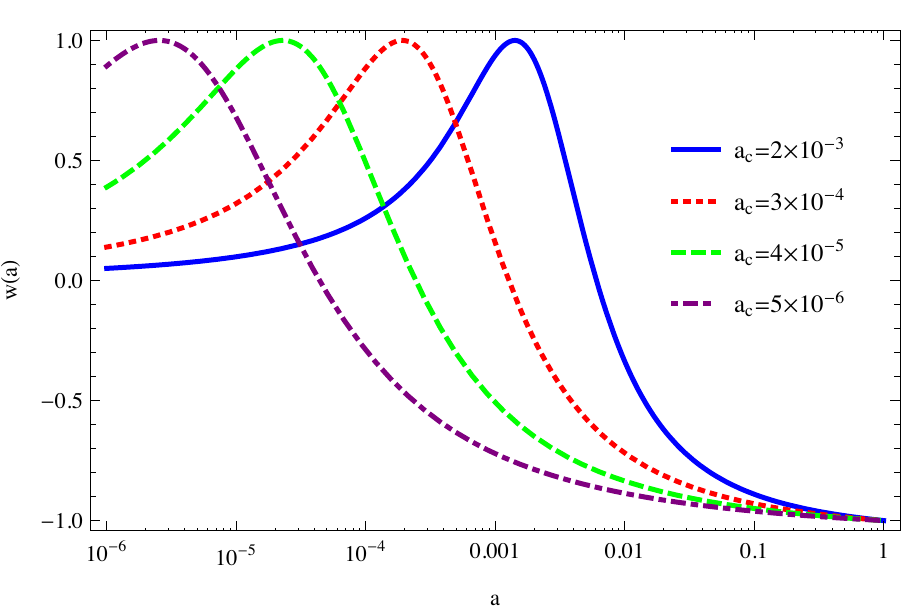}
\caption{Different parametrizations of the equation of state $w(a)$. The time of transition to dark energy behavior is determined by $a_c$. Initially it behaves like cold dark matter. It scales like $\Lambda$CDM for late $a$. The asymptotic value of $w(a)$ is variable in the MCMC.}
\label{fig:fig3}
\end{figure}

The background evolution of the energy density and equation of state of our EDE is illustrated in Figs.~\ref{fig:fig1},\,\ref{fig:fig1b},\,\ref{fig:fig2},\,\ref{fig:fig3}. Broadly, the dark energy is a fixed fraction of the dark matter density at early times.  It is interesting to note that in our model the ratio of dark energy to matter at early times can approach unity, as seen in Fig.~\ref{fig:fig1b}. Within the scope of our model, this suggests that dark energy could easily comprise half of the dark sector. The timing of the transition controls the abundance of dark energy at early times, which is seen as a peak in the EDE abundance in Fig.~\ref{fig:fig2}. For reasonable values of the parameters Fig.~\ref{fig:fig1b} shows that the peak can reach up to $1\%$ of the total energy density during last scattering, and thereby leave an imprint on the cosmic microwave background.

\subsection{Perturbations}
We assume that a scalar field underlies this model of dark energy, so that the small-scale speed of propagation of fluctuations in the dark energy is the speed of light, and fluctuations inside the horizon free stream away rather than cluster. The presence of early dark energy in the background still has an effect on the growth of perturbations in the dark matter and baryons \cite{Alam:2010rf}. Essentially, the presence of EDE suppresses the growth in the other components. There is a scale dependence to this phenomenon. Fluctuations of the dark matter and baryons that enter the horizon after the peak of EDE grow at a slower rate than modes that entered the horizon before the peak.

\begin{figure}[htbp]
	\includegraphics[width=\linewidth]{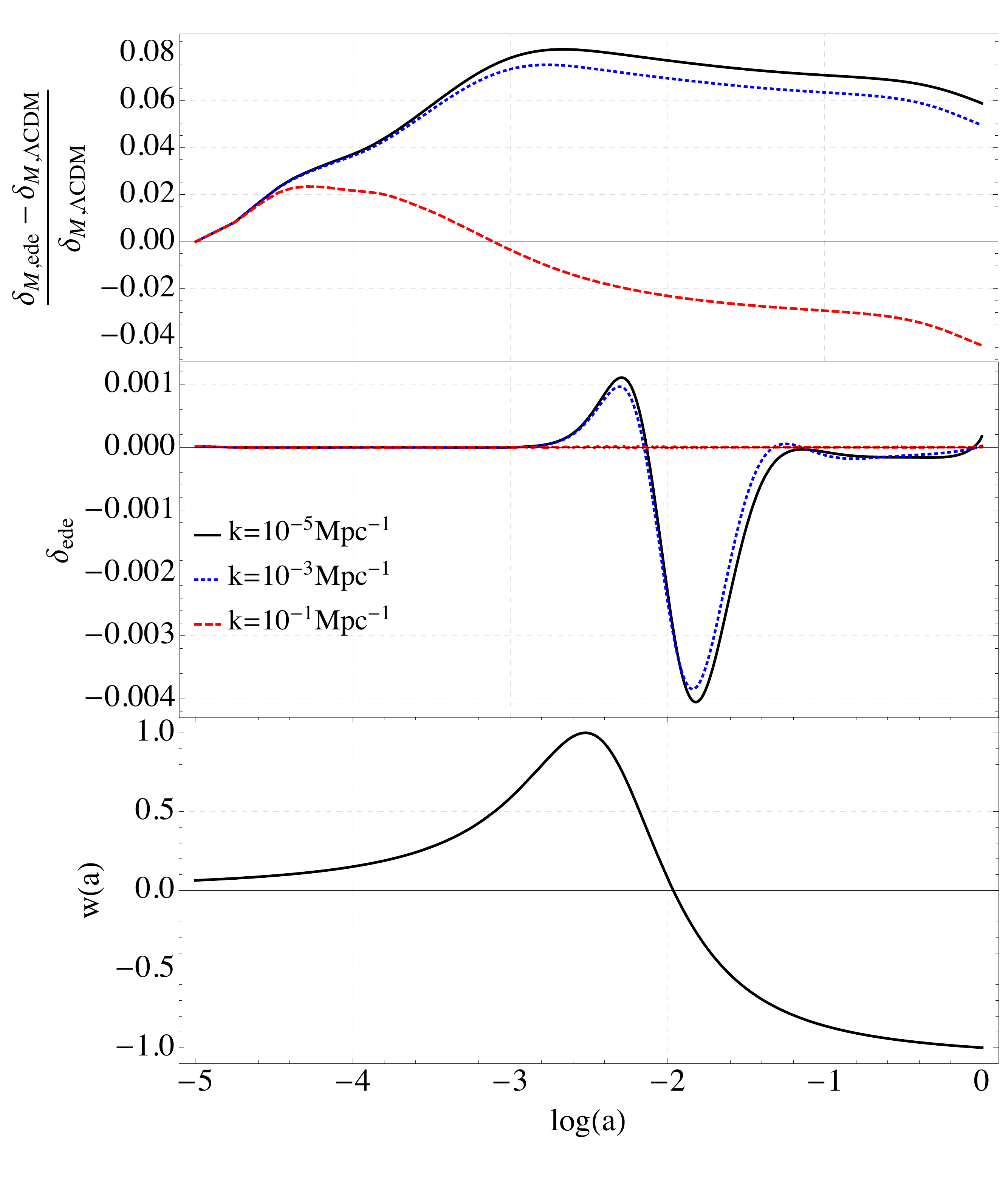}
	\caption{Perturbation behavior for $a_c = 0.004$. The top panel displays the relative deviations from the standard scenario in the synchronous gauge, based on a semi-analytic model. Initial conditions are equal for both models at $a=10^{-5}$. Large scale perturbations are most susceptible to early dark energy. \\ The middle panel shows the behavior of dark energy perturbations $\delta_{ede}=\delta\rho_{ede}/\rho_{ede}$ in the synchronous gauge. The bottom panel displays the equation of state.}
	\label{fig:perturbations}
\end{figure}

\begin{figure}[htbp]
\includegraphics[width=\linewidth]{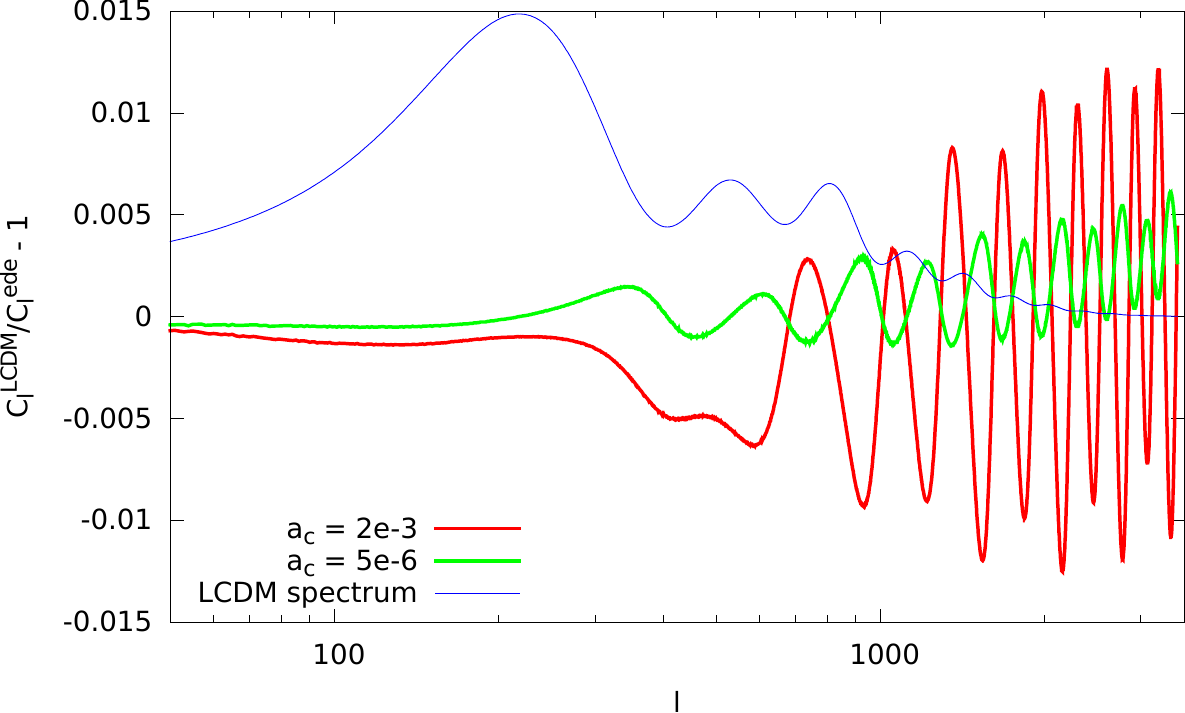}
\caption{$C_l^{\Lambda CDM}/C_l^{ede}-1$ vs. $l$. Equal normalizations at small $l$ lead to shifts at $l>600$. Increasing $a_c$ causes higher deviations from $\Lambda$CDM. The $\Lambda$CDM power spectrum is plotted in blue for comparison.}
\label{fig:Cls}
\end{figure}

We have compared the evolution of perturbations in our model to the standard $\Lambda$CDM case (as in Figs.~\ref{fig:perturbations},\,\ref{fig:Cls}). For the CMB we have adapted the Boltzmann code CAMB \cite{Lewis:1999bs}, based on CMBfast \cite{Seljak:1996is},  to calculate the CMB anisotropy spectrum and mass power spectrum for these EDE models. We note that CAMB already includes  perturbations of the dark energy, but assumes that the equation of state is a constant, with $w'=0$. So, the perturbation evolution equations in the file ``equations.f90'' need to be revised to include the $w'\neq 0$ term in the evolution of the momentum density perturbation.

To gain some insight into the behavior of the perturbations in the EDE we built a simple semi-analytic model that consists of perturbations of the dark matter and dark energy components including gravity as the system evolves from the radiation to the matter dominated epoch. We work in the synchronous gauge using evolution equations from Ref.~\cite{Ma:1995ey} and choose a cosmology with $a_c = 4\times 10^{-3}$ and $w_0 = -1$. This value of $a_c$ is large enough to make the difference with $\Lambda$CDM visible to the eye. The top panel shows the deviations of matter perturbations $\delta_M$ in the two models. Since we set up the evolution in each case with equal initial conditions at $a=10^{-5}$, then the deviations start to show when the universe has grown by an order of magnitude. One result from our analysis of the observational constraints, presented in the next section, is that deviations from $\Lambda$CDM can be as high as 5\% today. Short wavelength modes are affected the most, whereas for long wavelengths the discrepancy with the standard cosmological model drops significantly. 

The middle panel of Fig.~\ref{fig:perturbations} shows the perturbations in the dark energy component. Here absolute values of the perturbation $\delta_{\mathrm{EDE}}$ are given. These are small compared to the total matter perturbations, which can range up to $\delta_{\mathrm{M}}\sim0.15$ by the present day. The strongest perturbations in the EDE form right when the dark energy equation of state crosses $w=0$, as seen in Fig.~\ref{fig:perturbations}.  
  
In Fig.~\ref{fig:Cls} we present an example of the deviations expected in the CMB temperature power spectrum for our model relative to the predictions of the standard $\Lambda$CDM model. In general, we find that departures from the standard spectrum appear around $l\sim 600$ and become important at higher $l$'s.  Hence, it will be important to use data from a CMB experiment that extends beyond the range of WMAP \cite{Komatsu:2010fb} to small angular scales, such as the South Pole Telescope (SPT)  \cite{Keisler:2011aw, Reichardt:2011fv} or Planck \cite{Planck:2013kta}.

\section{Constraints}
\label{sec:constraints}

We modified CosmoMC \cite{Lewis:2002ah} for our equation-of-state model in order to constrain the parameters of our theory. Because this work was carried out prior to the release of the Planck data, we use WMAP \cite{Komatsu:2010fb} plus SPT \cite{Keisler:2011aw, Reichardt:2011fv} for CMB data. Hence we calculate the temperature power spectrum up to $l = 3000$. Apart from CMB data, we further constrained the parameters using SDSS DR7 matter power spectrum measurements,  the Union 2 supernova sample and HST data \cite{Reid:2009xm,Amanullah:2010vv,Riess:2009pu}.

The ten parameters that were varied in the MCMC are summarized in table \ref{tab:mcmcparams} where they are displayed with their most likely values and their 1D marginalized 68\% CL. The parameters $H_0$ and $\sigma_8$ are derived parameters which are calculated after each step in the chain. Technically $a_c$ is also a derived parameter, its relation to $1/x_c$ -- the corresponding MCMC parameter -- is obvious from $x_c = \ln(a_0/a_c)$. Using SPT required us to marginalize over the three foreground nuisance parameters that describe Poisson point sources, clustered point sources and added power from the kinetic SZ effect. These phenomena become important at small scales ($l \gtrsim 2000$) and therefore have to be included whenever we use SPT data.

\begin{center}
	\begin{table*}
\ra{1.3}
	  \begin{tabular}{l c c c c c c c c c c | c c}
	    \hline
	      & $\Omega_b h^2$ & $\Omega_{\mathrm{CDM}} h^2$  & $\theta$ & $\tau$ & $\Omega_K$ & $a_c $ & $Y_{He}$ & $w_0 $  & $n_s$ & $A_s$ & $H_0$ & $\sigma_8$ \\ \hline
	    center & 0.0224 & 0.1126 & 1.041 & 0.0858 & 0.0047 & 0 & 0.2478 & $-0.98$ & 0.969 & $2.26\times 10^{-9}$ & 70.2 & 0.796\\ 
	    min & 0.0218 & 0.1076 & 1.039 & 0.0672 & $4.03\times 10^{-4}$ & 0 & 0.2476 & $-1$ & 0.956 & $2.18\times 10^{-9}$ & 68.2 & 0.763\\ 
	    max & 0.0229 & 0.1178 & 1.043 & 0.1044 & 0.0096 & $4.8 \times 10^{-4}$ & 0.2481 & $-0.92$ & 0.982 & $2.34\times 10^{-9}$ & 72.3 & 0.829\\ \hline
	  \end{tabular}
	  \caption{MCMC parameter bounds: The minimum and maximum give the 1-$\sigma$ bounds.}
	  \label{tab:mcmcparams}
	\end{table*}
\end{center}

\begin{figure}[htbp]
	\includegraphics[width=\linewidth]{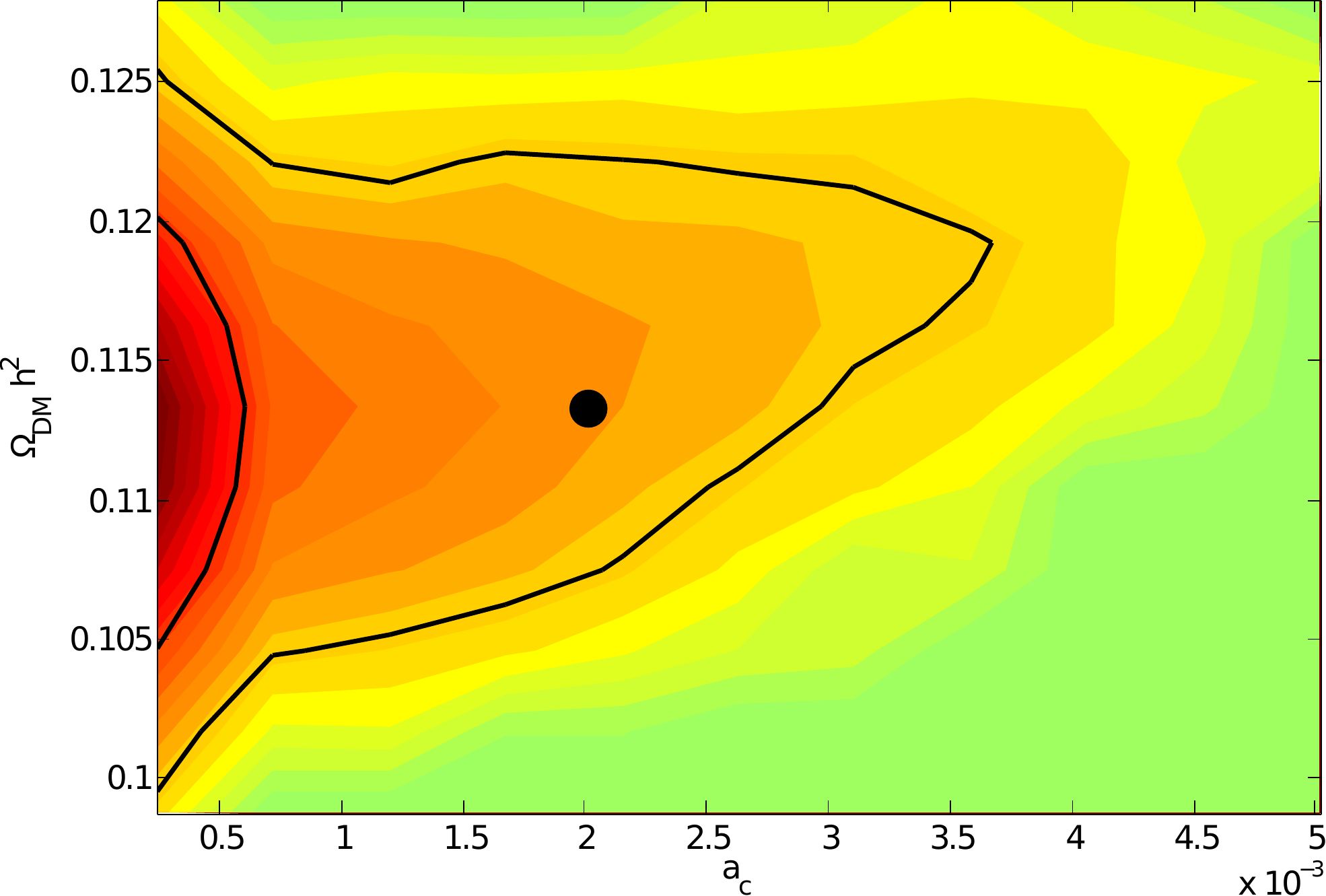}
	\caption{1- and 2-$\sigma$ contours for $a_c$ vs. $\Omega_{\mathrm{CDM}} h^2$. The black dot represents the value for the fiducial model in chapter \ref{sec:tt} with $a_c = 2\times 10^{-3}$ and $\Omega_{\mathrm{CDM}} h^2 = 0.1126$.}
	\label{fig:ac_Omh2}
\end{figure}

The results of the MCMC analysis are displayed most easily in terms of joint likelihood contour plots. In Fig.~\ref{fig:ac_Omh2} we have plotted the 1- and 2-$\sigma$ (68\% and 95\% CL) contours of $a_c$ vs. $\Omega_{\mathrm{CDM}} h^2$. We find the striking result that the parameter $a_c$ has an allowed range up to $a_c \sim 0.004$ ($z_c \sim 250$) -- well after last scattering. We placed a black dot in the parameter plane at $z_c=500$ and $\Omega_{\mathrm{CDM}} h^2 = 0.1126$ to provide a target for future observations. The black dot also refers to the fiducial model in our Planck TT forecast (Sec.~\ref{sec:tt}). We chose parameter values with a high overall likelihood, whereas the choice of the value for $a_c$ is less conservative but still within the 2-$\sigma$ contour.

\begin{figure}[htbp]
	\includegraphics[width=\linewidth]{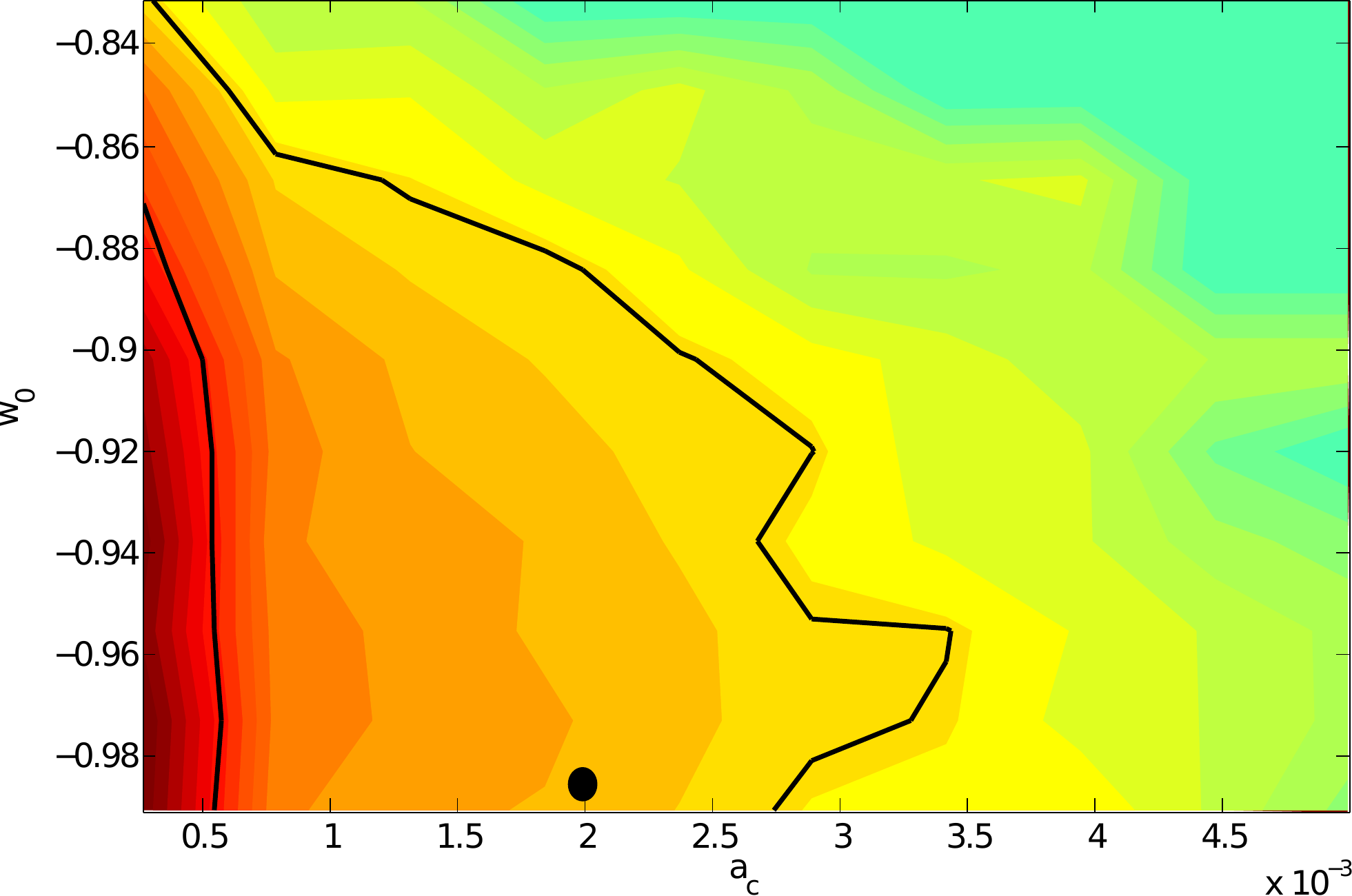}
	\caption{1- and 2-$\sigma$ contours for $a_c$ vs. $w_0$ today. Again, the black dot represents the value for the fiducial model in chapter \ref{sec:tt}.}
	\label{fig:ac_w0}
\end{figure}

The data constrains $w_0$, the asymptotic value of $w(a)$, similarly to models with a linearly varying equation of state. We applied a hard lower bound $w_0 \geq -1$. Not surprisingly a dark energy component that resembles a cosmological constant today allows for larger transition scale factors in Fig.~\ref{fig:ac_w0}.

\begin{figure}[htbp]
	\includegraphics[width=\linewidth]{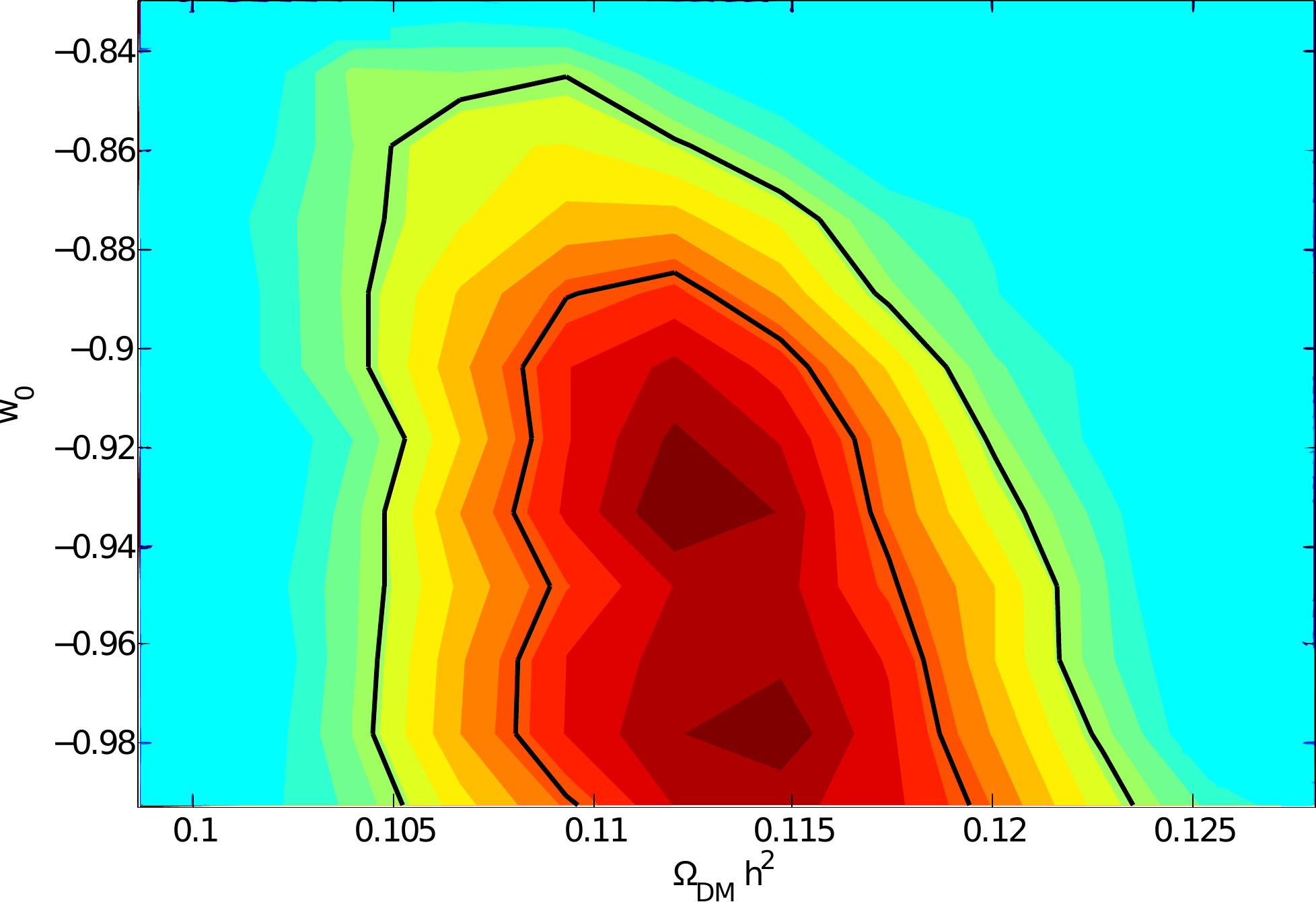}
	\caption{1- and 2-$\sigma$ contours for $\Omega_{\mathrm{CDM}} h^2$ vs. $w_0$ today.}
	\label{fig:Omh2_w0}
\end{figure}

The $\Omega_{\mathrm{CDM}} h^2$ vs. $w_0$ contours (Fig.~\ref{fig:Omh2_w0}) show similar results and simply reproduce results for a constant equation of state $w(a) = w_0$ with the typical banana shaped contours. The degeneracy seems to be less striking in our model, though.

\begin{figure}[htbp]
	\includegraphics[width=\linewidth]{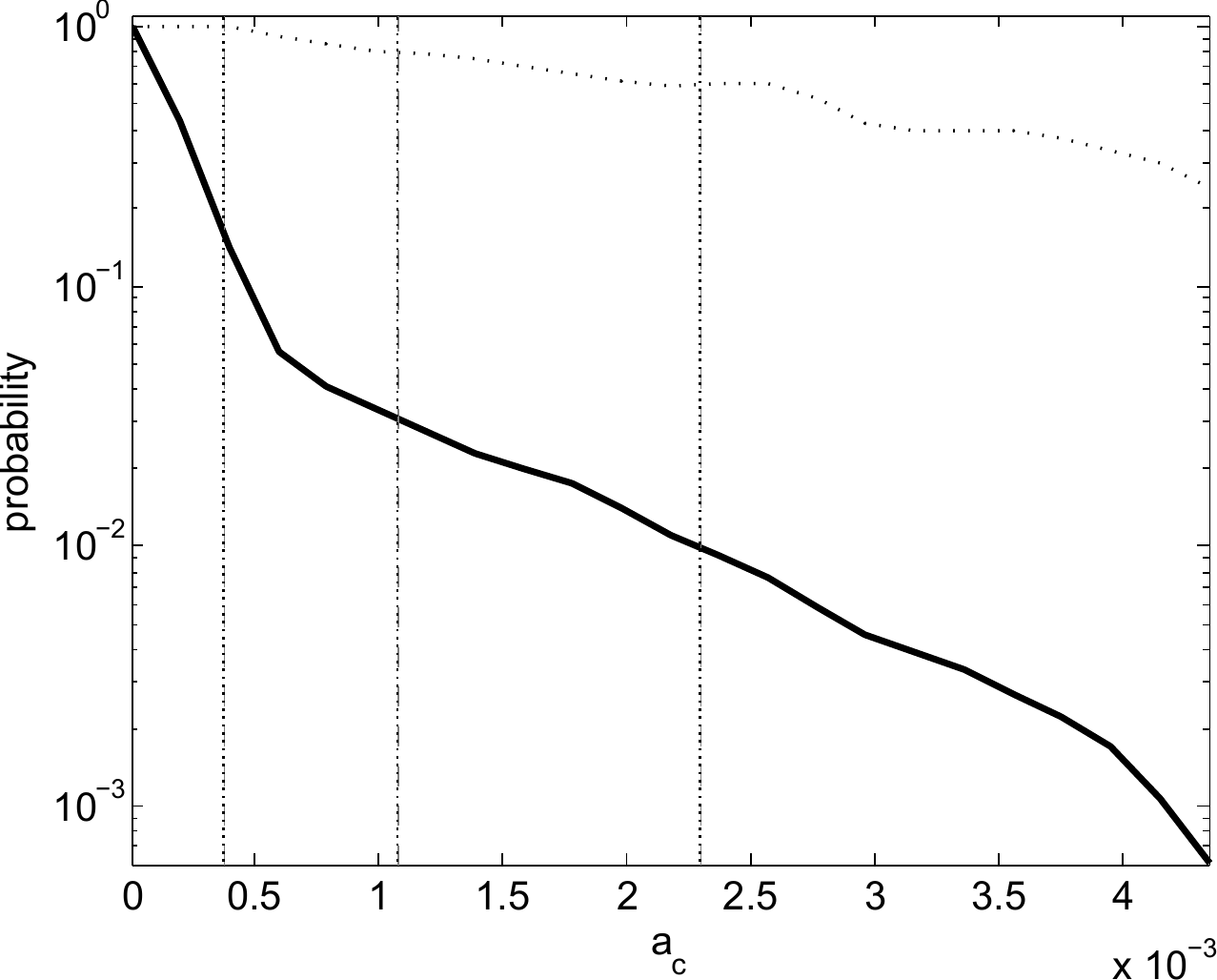}
	\caption{Probability distribution for $a_c$. The thin dotted line represents the likelihood. The grey dashed lines indicate the 1-, 2- and 3-$\sigma$ intervals for $a_c$.}
	\label{fig:ac_prob_logplot}
\end{figure}

The marginalized probability density (Fig.~\ref{fig:ac_prob_logplot}) for $a_c$ allows for freeze-out scales up to $a_c \sim 2.3\times 10^{-3}$ at 3-$\sigma$. The dotted line in the plot refers to the likelihood. The gray lines represent the 1- 2- and 3-$\sigma$ boundaries. 

\begin{figure}[t]
\includegraphics[width=\linewidth]{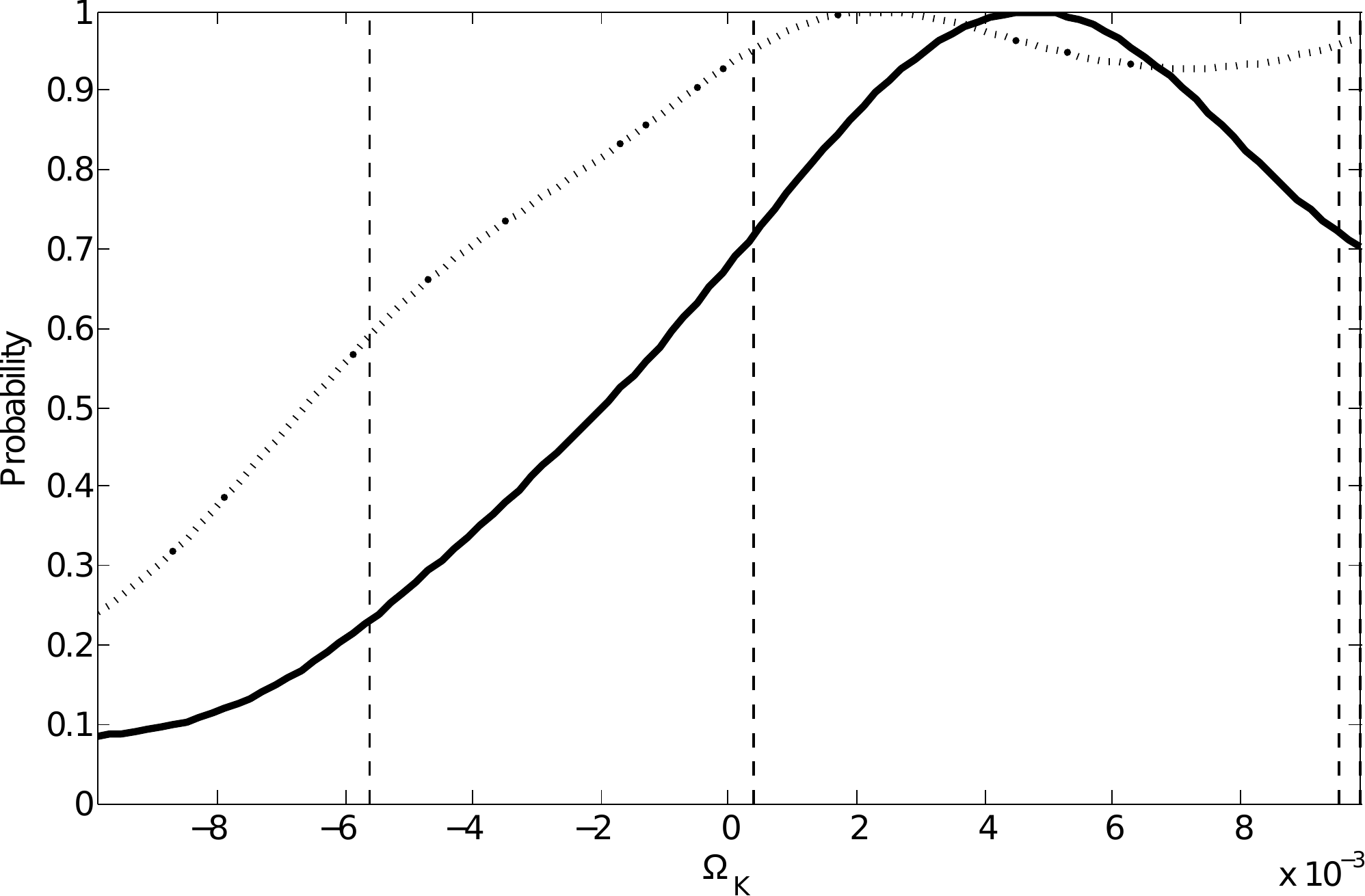}
\caption{$\Omega_{\mathrm{K}}$ probability distribution. The dotted line corresponds to the likelihood, the solid line represents the probability distribution. The dashed vertical lines depict the 1- and 2-$\sigma$ likelihoods.}
\label{fig:OmK}
\end{figure}

It is interesting to note that, while we restricted $\Omega_K$ to lie within $\pm 0.01$ and {\it a priori} expected the resulting distribution to have support well within those bounds, this did not turn out to be the case. Instead, the resulting MCMC probability distribution indicates that it should extend well beyond those scales. Moreover, it favors a closed universe although a flat model lies just outside the 1-$\sigma$ contour (see Fig.~\ref{fig:OmK}). 

In summary, the data weakly favors a $\Lambda$CDM cosmology, and allows deviations as far as $z_c \sim 400$ after marginalizing over all the other parameters. Indeed, the current, high quality data tightly constrains the physics of the recombination era, when trace levels of EDE may be present. And so we look ahead to future experiments to learn if there is any new information available with which to probe for EDE.

Recently Pettorino et. al. \cite{Pettorino:2013ia} examined the question what is the maximum amount of dark energy that is consistent with the data. Their constraints were obtained using CMB data only -- combined WMAP and SPT data probes the CMB up to multipoles $l \sim 3000$. Although they also used different models for the equation of state, they conclude, similar to us, that small scale CMB and weak lensing observations are important for future constraints to EDE.

\section{CMB Lensing Forecast}
\label{sec:lensing}
Dark energy not only varies the shape of the primary CMB power spectrum, but also influences large scale structure by enhancing or suppressing the growth of the matter power spectrum. This effect can be observed through the lensing of the CMB, which induces deflections on the primary CMB maps. With the lensed maps, one can reconstruct the lensing potential, which can be derived from the matter power spectrum. Even though EDE may have only played a role in the early universe ($z_c \sim 400$), it left a distinguishable imprint on the growth of structure at early times, that deviations from $\Lambda$CDM are still observable at small redshifts through the lensed maps \cite{Das:2012eb}.

With this in mind, we further explore the ability of futuristic CMB lensing experiments in constraining $a_c$, with the intent of understanding the limits in each kind of experiment and demonstrating the need to explore new phenomena in order to completely map out our cosmic history. To accomplish the first goal, we use the Fisher Matrix to estimate the 1-$\sigma$ errors of $a_c$ given some CMB lensing experiment input parameters, for the case of several different fiducial $a_c$ values. The advantages of using Fisher approximation compared to a MCMC analysis for this study are the low computation time costs for running multiple models and its computational tractability.

If one assumes the likelihood function is Gaussian, it can be written as
\begin{equation}
 \emph{L}(\boldsymbol{\theta}| \boldsymbol{d}) \propto \frac{1}{\sqrt{| \bar{C}(\boldsymbol{\theta})|}} \exp \left( -\frac{1}{2} \boldsymbol{d^{\dagger}}[\bar{C}(\boldsymbol{\theta})]^{-1} \boldsymbol{d}\right)
 \label{eqn:likelihood}
\end{equation}
where $\boldsymbol{d}$ is the data vector, $\boldsymbol{\theta}$ denotes the model parameters, and $\bar{C}$ is the covariance matrix of the modeled data. In our case,  $\boldsymbol{d}= \{a^T, a^E, a^d \}$ (where $T,\,E,\,d$ refer to temperature, E polarization, and weak lensing deflection), and $\boldsymbol{\theta}  = \{ \Omega_c h^2, \Omega_b h^2, \Omega_K, w, A_s, n_s, \tau, H_0, x_c \}$. Note that this is a smaller parameter space compared to the MCMC analysis -- including a parameter like $Y_{He}$ does not alter the observations on the trends of $a_c$.
\\*

Fisher information captures the curvatures of the Gaussian distributions around the fiducial values of the parameters. It can be written as

\begin{equation}
F_{ij} = - \frac{\partial^2 \log \textit{L}}{\partial \theta_i \partial \theta_j} \bigg|_{\boldsymbol{\theta} = \boldsymbol{\theta_0}}
\label{eqn:fishergen}
\end{equation}
where $\boldsymbol{\theta_0}$ contains the fiducial value of each parameter in the vector, which is chosen assuming that it maximizes the likelihood. For this study, we use the 2012 fiducial values $\Omega_c h^2 =  0.1123$,  $\Omega_b h^2  = 0.0226$, $\Omega_K = 0$, $w= -0.95$, $A_s = 2.46\times10^{-9}$, $n_s = 0.96$, $\tau = 0.084$, $H_0 = 70.4$, and we picked several values for $a_c$. We get the 1-$\sigma$ uncertainties of each parameter by marginalizing the rest, so for parameter $\theta_i$, 

\[
\sigma_i = (\sqrt{F^{-1}})_{ii} .
\] 
Given the form of the likelihood function \emph{L}, we can write

\begin{equation}
 F_{ij} = \sum_\textit{l} \sum_{XX', YY'} \frac{\partial C^{XX'}_\textit{l} }{\partial \theta_i}(Cov^{-1}_\textit{l})_{XX'YY'}  \frac{\partial C^{YY'}_\textit{l} }{\partial \theta_j}
 \end{equation}
where \textit{l} is the angular multipole of the power spectrum, $XX', YY' = \{TT, EE, TE, Td, dd, Ed\}$. The matrix $Cov_\textit{l}$ is the power spectrum covariance matrix at the \textit{l}-multipole and its form and the correlation coefficients are listed in Appendix \ref{app:fisherterms} \cite{Perotto:2006rj}.

Note that the power spectrum in the covariance matrix $\bar{C}^{XX'}_\textit{l}$ includes the Gaussian noise $N^{XX'}_\textit{l}$, where
\[
N^{XX'}_\textit{l}= s^2 \exp \left(l(l+1) \frac{\theta^2_{\textsc{fwhm}}}{8\log2}\right)
\] 
for XX' = \{TT, EE\}, $s$ is the instrumental noise in $\mu K$-radians, $\theta^2_\textsc{fwhm}$ is the full-width half-maximum beam size in radians. For XX' = dd, we use the quadratic estimator by Okamoto and Hu \cite{Okamoto:2003zw, futurCMB2009} to obtain the lensing reconstruction noise. The noise vanishes for $X \neq X'$, so that $N^{XX'} = 0$.

For this forecast, we looked into several experiments with different instrumental noise, beam size, sky coverage combinations and at a few $a_c$ values. The instrumental parameters are in the range of being an ideal experiment to one close to the next generation ground-based telescope or satellite \cite{EPICTeam2008, Bock:2009xw, KISS2012}. The fiducial values $a_c$ = 2.48$\times 10^{-3}$, 1.50$\times 10^{-3}$, 0.91$\times 10^{-3}$, and 0.55 $\times 10^{-3}$, are picked to be within 3-$\sigma$ of the current data constraints and relevant to detection.

We chose beam sizes of 1 arcminute and 5 arcminutes. Current ground-based polarization experiments like \textsc{SPTPol}\cite{Austermann:2012ga} and \textsc{ACTPol}\cite{Niemack:2010wz} have $\sim$1 arcmin beam, while \textsc{PolarBear}\cite{Errard:2010bn} has $\sim$5 arcminutes beam. The instrumental noises are 1, 5, and 10 $\mu$K-arcmin. Experiment like \textsc{POLAR Array} will achieve a sensitivity between 5-10 $\mu$K-arcmin. The two sky coverages $f_{sky}$ are 0.1 and 0.8, where the former is in accordance with ground based observations at the South Pole while the latter is what most satellites hope to achieve.

\begin{figure}[htbp]
	\includegraphics[width=\linewidth]{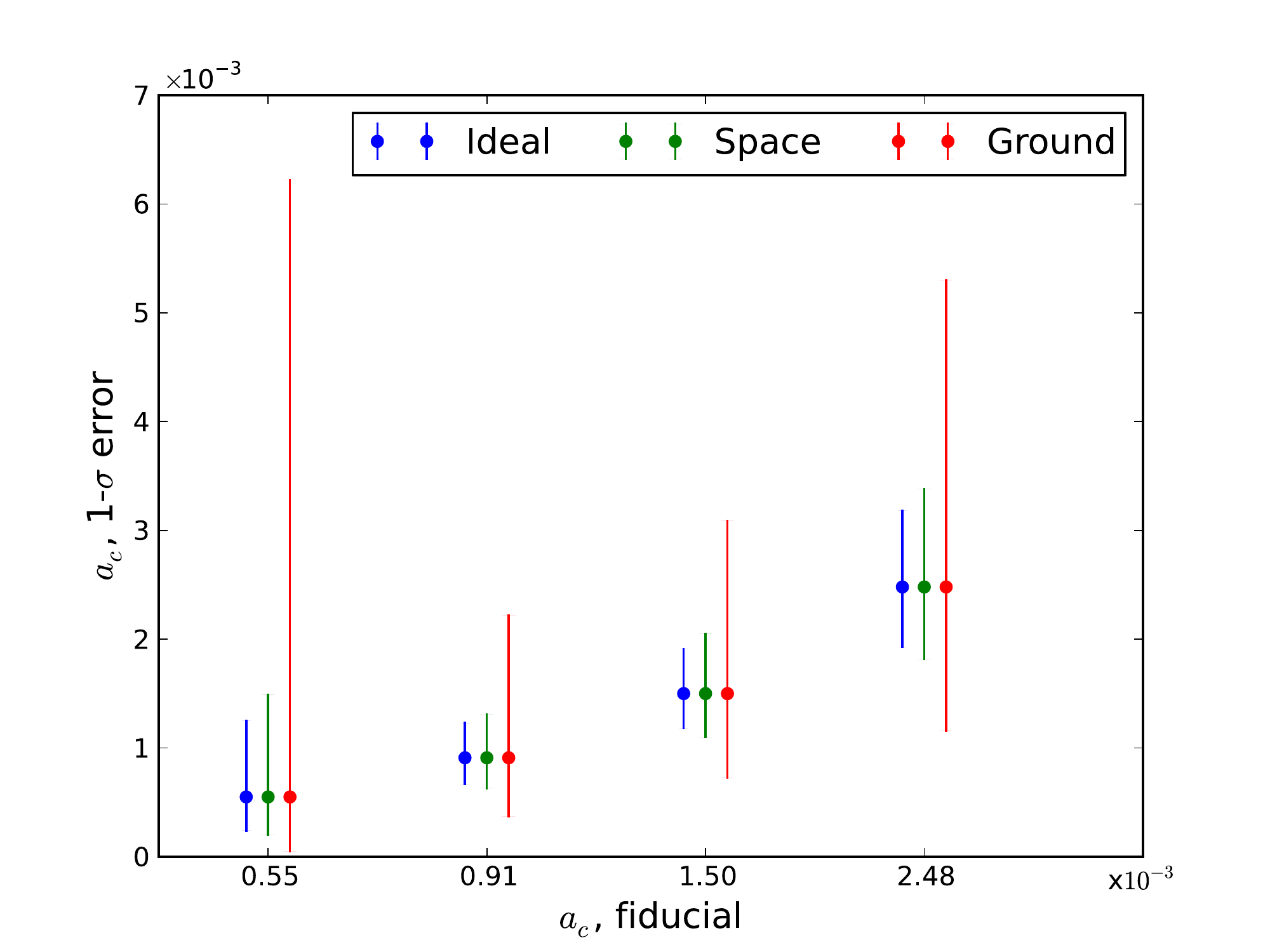}
	\caption{1-sigma constraints on $a_c$ for each fiducial $a_c$ values, shown by the error bars. The experiments exemplified here have the following inputs: \textit{Ideal} - 1$\mu$K-arcmin noise, 1' beam, $f_{sky}$=0.8;  \textit{Space} - 5$\mu$K-arcmin noise, 5' beam, $f_{sky}$=0.8; \textit{Ground} - 5$\mu$K-arcmin noise, 1' beam, $f_{sky}$=0.1}
	\label{fig:ideal_space_ground_constraint_ac}
\end{figure}

Figure \ref{fig:ideal_space_ground_constraint_ac} highlights how the constraints on $a_c$ change given the experimental setups specific to next generation ground based telescopes and space satellites as compared to an ideal experiment. The rest of the forecast results are in Tables \ref{table:lensingforecasts1} and \ref{table:lensingforecasts2} in Appendix \ref{app:fisherterms}. We observe that to constrain the transition redshift of this EDE model, a large sky coverage is essential in order to simultaneously constrain the low $l$ multipoles well. 

We see that for our EDE model, as long as $a_c$ is big enough, i.e., the dark energy transition happens at a late enough redshift, then we can have reasonable constraints on this parameter. However, once $a_c$ decreases to the point where the cosmology looks like  $\Lambda$CDM to the CMB and lensing, then distinguishing between different models will be a challenge. Given the model indicated with the black dot in Figs.~\ref{fig:ac_Omh2},\,\ref{fig:ac_w0} ($a_c = 2\times 10^{-3}$), then lensing data can easily distinguish this scenario from a $\Lambda$CDM background already at the 1-$\sigma$ level (see Table~\ref{table:lensingforecasts1}). Weak lensing puts tight constrains on the $a_c=2.48\times10^{-3}$ model that make the transition happen well after last scattering.

Tables \ref{table:lensingforecasts1} and \ref{table:lensingforecasts2} in Appendix \ref{app:forecast} list [$a_{c, low}-a_{c,fid}$, $a_{c, high}-a_{c,fid}$] for each $a_{c, fid}$ given an experimental setup, where $a_{c, low}$ and $a_{c, high}$ are values of $a_c$ 68\% C.L. from $a_{c,fid}$. They are asymmetric because they are derived from 1-$\sigma$ constraints on $x_c$ marginalizing the rest of the parameters. We see that as we increase the sky coverage $f_{sky}$ the constraints improve by $1/\sqrt{f_{sky}}$ as expected. With better sensitivity (lower noise), the constraints also improve and it is interesting to note that the constraints improve more for the 5 arcmin case than the 1 arcmin case. On the other hand, five times smaller beam size does not gain five times better constraints and the constraints are within 10\% of each other between the 1 arcminute and the 5 arcminute beam.

\section{Planck TT forecasts}
\label{sec:tt}

To compute the likelihood contour forecasts in anti\-cipation of an analysis of the Planck data \cite{Ade:2013zuv}, we perform a Fisher-matrix based analysis. In this case the Fisher matrix $F_{ij}$ (\ref{eqn:fishergen}) takes the simple form \cite{Dodelson:2003ft}
\begin{equation*}
	F_{ij}=\sum_l \frac{1}{(\delta C_l )^2}\frac{\partial C_l}{\partial \theta_i}\frac{\partial C_l}{\partial \theta_j} \bigg|_{\boldsymbol{\theta_0}}   .
\end{equation*}
The summation runs over all values of $l$ for which we expect to get measurements. For large $l$ the terms in the sum will be suppressed by large values of $\delta C_l$. Because the Fisher matrix and therefore the entire forecast depends crucially on the fiducial parameters chosen, we chose the parameter set that produces the maximum likelihood (see Table \ref{table:forecasts}).

The errors $\delta C_l$ include the effects of cosmic variance, incomplete sky coverage and detector noise. Apart from the sky coverage ratio $f_{sky}$ one needs the FWHM beam size $\theta_\mathrm{FWHM}$ and the sensitivity per pixel $\tau_\mathrm{pix}$ in $\mu$K. The beam size is calculated as $\sigma \simeq 0.00742 \times \theta_\mathrm{FWHM} ^2$ for small beam sizes and $\theta_\mathrm{FWHM}$ in degrees. One defines an inverse weight $w^{-1} = (\theta_\mathrm{FWHM} \tau_\mathrm{pix})^2$ where $\theta_\mathrm{FWHM}$ enters in units of radians. These variables enter the error formula in the following way \cite{Eisenstein:1998tu}:
\begin{equation*}
	\delta C_l = \sqrt{\frac{2}{(2l+1)f_{sky}}} \left( C_l + w^{-1} e^{l^2\sigma ^2} \right)   .
\end{equation*}
For Planck, we use $\theta_\mathrm{FWHM} = 7.1$ arcminutes and $\tau_\mathrm{pix} = 2.2\,\mu\mathrm{K}$ which are the values of the 143 GHz high frequency instrument \cite{PlanckHFICoreTeam:2011az}. We assume a sky coverage of 65\% which is a conservative estimate for Planck.

\begin{figure}[htbp]
	\includegraphics[width=\linewidth]{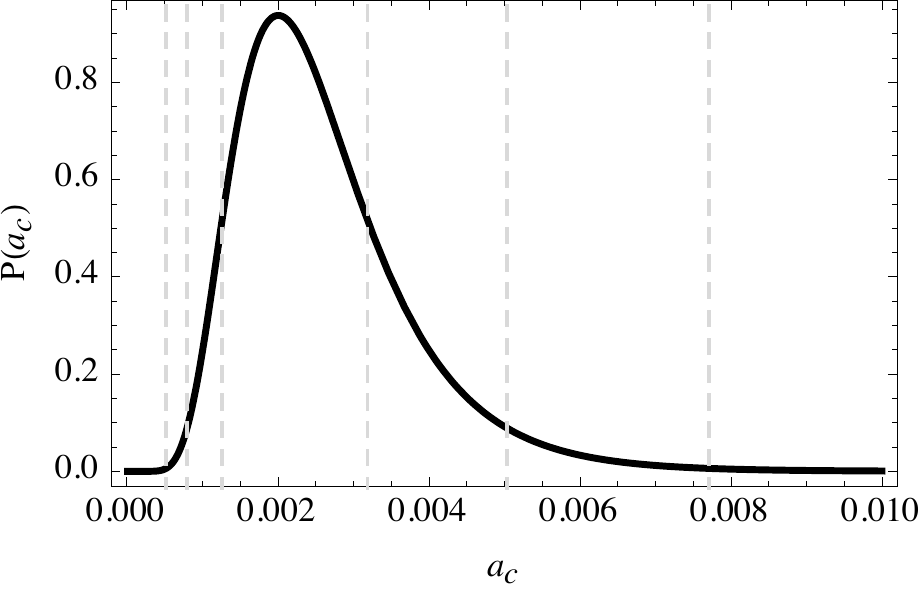}
	\caption{Predicted probability distribution for $a_c$. The grey dashed lines indicate the 1-, 2- and 3-$\sigma$ intervals for $a_c$. The mean value is at $a_c = 0.0021$. The distribution is not Gaussian since $a_c$ is a derived parameter in our model.}
	\label{fig:ac_pred}
\end{figure}

\begin{figure}[htbp]
	\includegraphics[width=0.9\linewidth]{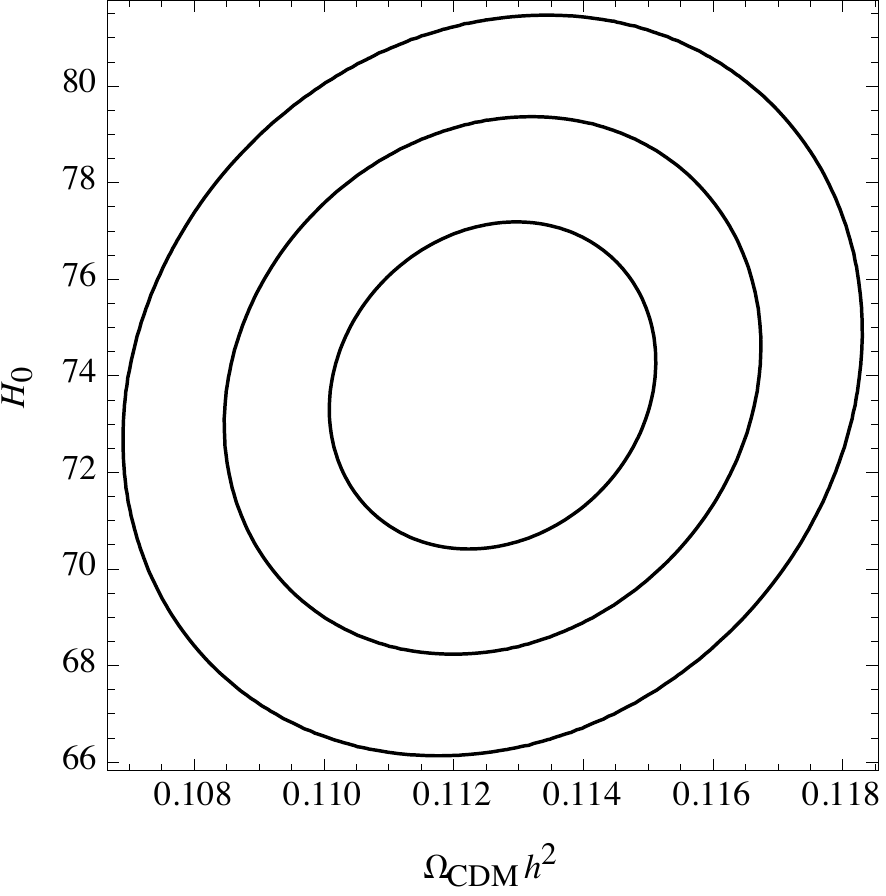}
	\caption{Predicted constraints for $\Omega_{CDM} h^2$ vs. the Hubble constant $H_0$. The fiducial model is centered around current WMAP constraints: $\Omega_{CDM}^{fiducial}h^2=0.1126$ and $H_0 =73.8\text{ km s}^{-1}\text{Mpc}^{-1}$. The contours are at 1-, 2- and 3-$\sigma$.}
	\label{fig:H0_Om_pred}
\vspace{.3cm}
	\includegraphics[width=\linewidth]{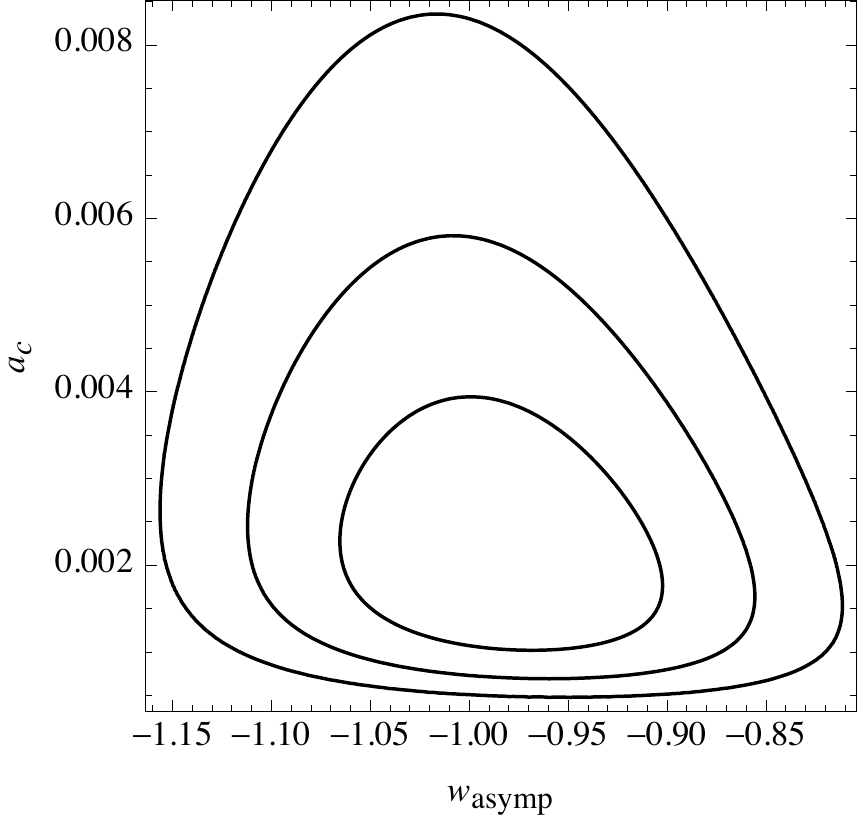}
	\caption{Predicted constraints for $\omega_\text{asymp}$ vs. $a_c$. The fiducial model is centered around $\omega_\text{asymp}=-0.984$ and $a_c =0.0021$. The contours are at 1-, 2- and 3-$\sigma$.}
	\label{fig:w0_ac}
\end{figure}

We further restrict the forecasted parameters by applying a prior on $H_0$ with $\sigma_{H_0} = 2.4\, {\rm km\, s}^{-1} {\rm Mpc}^{-1}$. The results of the forecasts are summarized in Figs.~ \ref{fig:ac_pred}, \ref{fig:H0_Om_pred}, \ref{fig:w0_ac}.

The $a_c$ contours are not elliptical because it is a derived parameter: We analyzed our model varying $x_c = \ln(a_0/a_c)$. Here we are plotting the contours for $a_c$ itself. Therefore the contours get deformed after computing the forecast of the model parameter. The fiducial parameters and standard deviations $\sigma_{\mathrm{forecast}}$ for the forecast are given in Table \ref{table:forecasts}. The improvement in the constraint on $a_c$ is comparable to the expected improvement in the constraints on conventional parameters such as $\Omega_{CDM} h^2$.  For the case of the fiducial model indicated by the black circles in Figs.~\ref{fig:ac_Omh2},\,\ref{fig:ac_w0} we forecast that Planck should be able to reject the $\Lambda$CDM model at the 3-$\sigma$ level.

\begin{center}
	\begin{table*}
	\ra{1.3}
\begin{tabular}{@{}lclclclclclclclclclclcl@{}}
	    \hline
	      & $\Omega_b h^2$ & $\Omega_{CDM} h^2$  & $\Omega_K$ & $Y_{He} $ & $\tau$ && $n_s$ & $A_s$ & $w_0 $ && $H_0$ && $a_c$ \\ \hline
	    fiducial value & 0.02255 & 0.1126 & 0 & 0.247779 & 0.088 && 0.968 & $2.2551 \times 10^{-9}$ & $-0.984$ && 73.8 && 0.002 \\ 
	    $\sigma_{\mathrm{forecast}}$ & $7.4\times 10^{-5}$ & 0.00145 & 0.0008 & - & 0.00193 && 0.0036 & - & 0.05 && 2.24 && 0.0007 \\
	    \hline
	  \end{tabular}
	  \caption{Parameters without standard deviations are marginalized over in the analysis.}
	  \label{table:forecasts}
	\end{table*}
\end{center}

\section{Outlook and Summary}

We have proposed a new EDE scenario in which dark energy tracks the dark matter until a transition whereupon the dark energy decouples and eventually becomes potential energy dominated. The equation of state trajectory features a spike at this transition as $w$ reaches $+1$ before dropping down to $w\rightarrow -1$. According to this model the dark energy could be the dominant component of a dark sector at early times. The value of $\rho_{\rm EDE}/\rho_M$ can be close to unity at early times (see Fig.~\ref{fig:fig1b}).

We performed an MCMC parameter search and found that the transition could be as late as $z_c \sim 250$ depending on the value of the parameter $\Omega_{CDM} h^2$. After marginalizing over all other parameters we find a 3-$\sigma$ bound of $z_c \gtrsim 400$. Hence, despite recent claims for a hint of excess relativistic degrees of freedom in the cosmological fluid \cite{Hamann:2010bk,Hou:2011ec,Smith:2011es,Archidiacono:2011gq} -- which seem inconsistent with recent Planck results \cite{Ade:2013zuv} -- we do not find positive evidence for a spike in the equation of state at early times. Future CMB temperature and weak lensing measurements may put tighter constraints on the possibility of such a transition. Beyond the CMB we expect a wide range of cosmological experiments will be able to test the viability of this model.

The physics of the 21cm transition will provide a complementary data set to weak lensing. Probes like the Square Kilometer Array will be able to detect gravitational lensing of the 21 cm transition of neutral hydrogen up to $z\sim 6$ \cite{Metcalf:2008gq}, whereas conventional lensing constrains the matter power function up to $z\sim1.5$. The SKA will therefore have the ability to trace the evolution of the HI mass function over 75\% of cosmic time and would give us new insights into dark matter halo mass functions. With those properties the SKA would become the main instrument for measuring EDE with BAOs.

Our next step is to analyze the model with respect to the Planck data \cite{Ade:2013zuv}, to see if it matches up to our expectations, and whether the results differ in any significant way from the WMAP+SPT analysis performed herein. Another task which we leave for the future is to determine if a more sophisticated model can be built that describes a dark energy species that decouples from dark matter.
 
\acknowledgments
This research was carried out in part at the Jet Propulsion Laboratory, run by the California Institute of Technology under a contract from NASA, and Dartmouth College and was funded through the JPL Strategic University Research Partnership (SURP) Program. Part of this work was supported by the Keck Institute of Space Studies and we thank colleagues at the ``CMB Polarization Cosmology in the Coming Decade" for stimulating discussions.
 
\vfill
\eject

\onecolumngrid
\appendix

\section{CMB lensing forecast results}\label{app:forecast}
The derived 1-sigma error of $a_c$ from the Fisher forecast over the whole grid of experimental noise = {1,5,10} $\mu $K-arcmin, beamsizes = 1, 5 arcminutes, sky coverage $f_{sky}$ = 0.1 and 0.8.

\begin{table*}[!h]\centering
\ra{1.3}
\begin{tabular}{@{}lclclclclcl@{}}\hline
&& {$a_c = 0.55  \times 10^{-3}$} & \phantom{abc}&{$a_c = 0.91 \times 10^{-3}$} & 
\phantom{abc} &{$a_c = 1.50 \times 10^{-3}$} & \phantom{abc} & {$a_c = 2.48 \times 10^{-3}$}\\ \hline
$\mathbf{f_{sky}=0.8}$\\
1($\mu $K-arcmin) &&  -0.31, 0.70  && -0.24, 0.32 && -0.32, 0.41&& -0.55, 0.70 \\
5($\mu $K-arcmin) && -0.32, 0.75 && -0.25, 0.34 && -0.33, 0.43  && -0.59, 0.77 \\
10($\mu $K-arcmin) &&  -0.32, 0.79   && -0.25, 0.35  && -0.36, 0.47   && -0.61, 0.80 \\
$\mathbf{f_{sky}=0.1}$\\
1($\mu $K-arcmin) &&  -0.05, 5.07  && -0.52, 1.23 && -0.73, 1.43 && -1.25, 2.51 \\
5($\mu $K-arcmin) &&  -0.50, 5.67 && -0.54, 1.31 && -0.77, 1.59  && -1.32, 2.82 \\
10($\mu $K-arcmin) && -0.51, 6.18  && -0.55, 1.38  && -0.80, 1.71   && -1.38, 3.09 \\
\hline
\end{tabular}
\caption{Derived 1-$\sigma$ constraints on the indicated fiducial $a_c$. All experiments have 1' beam. The entries are [$a_{c, low}$-$a_{c,fid}$, $a_{c, high}$-$a_{c,fid}$] respectively, where $a_{c, low}$ and $a_{c, high}$ are values of $a_c$ 1-$\sigma$ from the fiducial.}
\label{table:lensingforecasts1}
\end{table*}

\begin{table*}[!h]\centering
\ra{1.3}
\begin{tabular}{@{}lclclclclcl@{}}\hline
&& {$a_c = 0.55  \times 10^{-3}$} & \phantom{abc}&{$a_c = 0.91 \times 10^{-3}$} & 
\phantom{abc} &{$a_c = 1.50 \times 10^{-3}$} & \phantom{abc} & {$a_c = 2.48 \times 10^{-3}$}\\ \hline
$\mathbf{f_{sky}=0.8}$\\
1($\mu $K-arcmin) && -0.33, 0.80 && -0.25, 0.35 && -0.36, 0.47 && -0.59, 0.77 \\
5($\mu $K-arcmin )&& -0.35, 0.94 && -0.28, 0.40 && -0.40, 0.55  && -0.66, 0.90 \\
10($\mu $K-arcmin) && -0.36, 1.03 && -0.29, 0.43 && -0.43, 0.61 && -0.71, 1.00 \\
$\mathbf{f_{sky}=0.1}$\\
1($\mu $K-arcmin) &&  -0.51, 6.32  && -0.55, 1.39 && -0.79, 1.68 && -1.34, 2.93 \\
5($\mu $K-arcmin) &&  -0.52, 8.54 && -0.59, 1.64 && -0.87, 2.09  && -1.45, 3.50 \\
10($\mu $K-arcmin) && -0.52, 10.34  && -0.61, 1.83  && -0.92, 2.38   && -1.52, 3.93 \\
\hline
\end{tabular}
\caption{Derived 1-$\sigma$ constraints on the indicated fiducial $a_c$. All experiments have 5' beam. The entries are [$a_{c, low}$-$a_{c,fid}$, $a_{c, high}$-$a_{c,fid}$] respectively, where $a_{c, low}$ and $a_{c, high}$ are values of $a_c$ 1-$\sigma$ from the fiducial.}
\label{table:lensingforecasts2}
\end{table*}

\section{Fisher matrix power spectrum covariance terms}\label{app:fisherterms}
Here are the terms in the Fisher matrix formulation. 

\[ Cov_\textit{l} = \frac{2}{(2\textit{l}+1) f_{sky}} \left( \begin{array}{cccccc} 
\Xi_{TTTT} & \Xi_{TTEE}  & \Xi_{TTTE} &  \Xi_{TTTd} & \Xi_{TTdd}  & \Xi_{TTEd} \\
\Xi_{TTEE} & \Xi_{EEEE} & \Xi_{TEEE} & \Xi_{TdEE} & \Xi_{EEdd} & \Xi_{EEEd} \\
\Xi_{TTTE} & \Xi_{TEEE} & \Xi_{TETE} &  \Xi_{TETd} & \Xi_{TEdd} & \Xi_{TEEd}\\
\Xi_{TTTd} & \Xi_{TdEE} & \Xi_{TETd} & \Xi_{TdTd} & \Xi_{Tddd} & \Xi_{TdEd} \\
\Xi_{TTdd} & \Xi_{EEdd} & \Xi_{TEdd} & \Xi_{Tddd} & \Xi_{dddd} & \Xi_{ddEd} \\
\Xi_{TTEd} & \Xi_{EEEd} & \Xi_{TEEd} & \Xi_{TdEd} & \Xi_{ddEd} & \Xi_{EdEd} \end{array} \right)\]

\bigskip
\begin{eqnarray*}
\Xi_{TTTT} = (\bar{C}_l^{TT})^2& \hspace{30 mm}& \Xi_{TTEE}  = (\bar{C}_l^{TE})^2  \\
\Xi_{TTTE}  =  \bar{C}_l^{TT}  \bar{C}_l^{TE}&   \hspace{30 mm}& \Xi_{TTTd}  =  \bar{C}_l^{TT}  \bar{C}_l^{Td}\\
\Xi_{TTdd}   =  (\bar{C}_l^{Td})^2  & \hspace{30 mm}& \Xi_{TTEd}   =  \bar{C}_l^{TE}  \bar{C}_l^{Td} \\
\Xi_{EEEE} =  (\bar{C}_l^{EE})^2  &  \hspace{30mm}& \Xi_{TEEE} =  \bar{C}_l^{EE}  \bar{C}_l^{TE}  \\
\Xi_{TdEE} =  \bar{C}_l^{TE}  \bar{C}_l^{Ed}   & \hspace{30 mm}& \Xi_{EEdd} =  \bar{C}_l^{Ed}  \bar{C}_l^{Ed}  \\
\Xi_{EEEd} =  \bar{C}_l^{EE}  \bar{C}_l^{Ed}  &  \hspace{30 mm}& \Xi_{TETE} =  \frac{1}{2} \left[(\bar{C}_l^{TE})^2 + \bar{C}_l^{TT} \bar{C}_l^{EE} \right] \\
\Xi_{TETd} =  \frac{1}{2} \left[\bar{C}_l^{TT} \bar{C}_l^{Ed} + \bar{C}_l^{TE} \bar{C}_l^{Td} \right]  &  \hspace{30 mm}& \Xi_{TEdd} =   \bar{C}_l^{Td}  \bar{C}_l^{Ed}  \\
\Xi_{TEEd} =   \frac{1}{2} \left[\bar{C}_l^{EE} \bar{C}_l^{Td} + \bar{C}_l^{TE} \bar{C}_l^{Ed} \right]  & \hspace{30 mm}& \Xi_{TdTd} =  \frac{1}{2} \left[(\bar{C}_l^{Td})^2 + \bar{C}_l^{TT} \bar{C}_l^{dd} \right] \\
 \Xi_{Tddd} =  \bar{C}_l^{Td}  \bar{C}_l^{dd} &  \hspace{30 mm}& \Xi_{TdEd} =   \frac{1}{2} \left[\bar{C}_l^{TE} \bar{C}_l^{dd}+ \bar{C}_l^{Td} \bar{C}_l^{Ed} \right] \\
\Xi_{dddd} =  (\bar{C}_l^{dd})^2  & \hspace{30 mm}& \Xi_{ddEd} =  \bar{C}_l^{dd}  \bar{C}_l^{Ed}  \\
\Xi_{EdEd} = \frac{1}{2} \left[\bar{C}_l^{EE} \bar{C}_l^{dd} + (\bar{C}_l^{Ed})^2 \right] 
\end{eqnarray*}

\twocolumngrid

\end{document}